\documentclass[aps,prep,epsfile,nofootinbib]{revtex4}
\usepackage{graphics}
\usepackage{slashed}
\usepackage{amssymb}
\usepackage{lscape}
\usepackage{amsmath}
\usepackage{graphicx}
\graphicspath{ {images/} }
\begin{document}

\title{Space-time waves from a collapsing universe with a gravitational attractor}

\author{$^{1}$ Jaime Mendoza Hern\'andez\footnote{E-mail: jaime.mendoza@alumno.udg.mx}, $^{2,3}$ Mauricio Bellini
\footnote{{\bf Corresponding author}: mbellini@mdp.edu.ar}, $^{1}$ Claudia Moreno\footnote{E-mail: claudia.moreno@cucei.udg.mx} }
\address{$^1$ Departamento de F\'{\i}sica, Centro Universitario de Ciencias Exactas e Ingenier\'{\i}as, Universidad de Guadalajara
Av. Revoluci\'on 1500, Colonia Ol\'impica C.P. 44430, Guadalajara, Jalisco, M\'exico. \\
$^2$ Departamento de F\'{\i}sica, Facultad de Ciencias Exactas y
Naturales, Universidad Nacional de Mar del Plata, Funes 3350, C.P. 7600, Mar del Plata, Argentina.\\
$^3$ Instituto de Investigaciones F\'{\i}sicas de Mar del Plata (IFIMAR), \\
Consejo Nacional de Investigaciones Cient\'ificas y T\'ecnicas
(CONICET), Mar del Plata, Argentina.}
\begin{abstract}
We study a collapsing system attracted by a spherically symmetric gravitational source, with an increasing mass, that generates back-reaction effects that are the source of space-time waves. As an example, we consider an exponential collapse and the space-time waves emitted during this collapse due to the back-reaction effects, originated by geometrical deformation driven by the increment of the gravitational attracting mass during the collapse.
\end{abstract}
\maketitle

\section{Introduction}\label{1}

It is known that the universe is globally isotropic and homogeneous at cosmological scales\cite{1}. Today, these scales are in the range of $10^{8} - 10^{10} \; {\rm yl} $. However, at astrophysical scales the universe becomes inhomogeneous, and the expansion is $r$-dependent. The mass spectrum on astrophysical scales has a negative index, which decreases with the observed scale. The study of gravitational collapse is a very important topic that help us to understand many aspects of the universe on astrophysical scales. The seminal work of Oppenheimer and Snyder\cite{op} explored the gravitational collapse of a dust cloud. Furthermore, in that years Tolman and Blondi studied the collapse of a spherically symmetric and inhomogeneous distribution of dust\cite{t,b}. In 1969 R. Penrose proposed the cosmic censorship hypothesis (CCH)\cite{pen} where he stated that any cosmological singularity will always be covered by an event horizon. The statement of this conjecture can be treated as weak and strong versions\cite{w,s}. More recently, a collapsing system described by a fluid with a heat flux has been treated in \cite{Sharma}. Furthermore, the dynamics of a gravitational collapse where  dissipation is included, was studied in\cite{HS} and the collapse driven by a scalar field that can avoid the final singularity \cite{B,J}
and a dust collapse in Einstein-Gauss-Bonnet gravity was explored in \cite{mal}. The critical collapse of a massless scalar field in a semiclassical treatment of loop quantum gravity was recently considered in \cite{pul}. An important case for cosmology is the study of a cosmological collapse driven by a scalar field, where the spherical symmetry is preserved\cite{Gundlach,GJ,G}. In this framework, a model for a dynamical collapse driven by a scalar field, when a relativistic observer falls co-moving with the collapse and cross the horizon of a Schwarzschild black-hole (BH), but space-time preserves hyperbolicity, was studied in\cite{jmc}. This system was demonstrated that can avoid the final singularity\cite{jmc1}.

In this work we extend these previous analysis in order to study the collapse originated by a gravitational attractor in an exponential collapsing universe. We explore this topic by considering a gravitational attractor on a background homogeneous and isotropic universe in an exponential collapse. The idea consists to consider the attractor in the boundary conditions of the minimum action principle as a flux (due to the gravitational field): $\delta \Phi= \lambda(r,t)\,g^{\alpha\beta}\,\delta g_{\alpha\beta}$. This attracting flux brokes the homogeneity of the system to incorporate a negative cosmological parameter $\lambda(r,t)<0$, in the new Einstein's equations. The manuscript is organized as follows: in Sect. II we revisit and extend the study of Relativistic Quantum Geometry (RQG) for a flux originated by an increasing gravitational source, and then inhomogeneous, such that the cosmological parameter that appears in the Einstein equations becomes $r,t$-dependent. In Sect. III we describe the background dynamics for a model that describes a collapsing system, with a gravitational attractor which has an increasing mass. In Sect. IV we develop a particular case of the model introduced in Sect. III. In particular, we describe back-reaction effects described by the dynamics of the geometric scalar field $\sigma$, which constitute the source for the space-time waves. Finally, in Sect. V we develop some final comments and conclusions.

\section{RQG and $\lambda(r,t)$ cosmological parameter}\label{2}

We consider the Einstein-Hilbert action ${\cal I}$, which describes gravitation and matter
\begin{equation}\label{act}
{\cal I} =\int_V d^4x \,\sqrt{-g} \left[ \frac{R}{2\kappa} + {\cal L}_m\right],
\end{equation}
where $\kappa = 8 \pi G$, ${{\cal L}_m}$ is the Lagrangian density that describes the background physical dynamics and
$R$ is the background scalar curvature.

We are aimed to study the flux $\delta \Phi$ originated by a gravitational source, after considering the variation of the Einstein-Hilbert action with boundary terms included. In this work we shall consider the case where
$\delta R_{\alpha\beta}$ is related to the variation of the metric tensor
\begin{equation}\label{f2}
\delta R_{\alpha\beta}= \lambda(r,t)\,\delta g_{\alpha\beta},
\end{equation}
where $\lambda(r,t)$ is the called cosmological parameter. However, in general this parameter can be a function of all the coordinates: $\lambda\equiv \lambda(x)$. Using the fact that $\delta \left[g_{\alpha\beta}\, g^{\alpha\beta}\right]=0$, we obtain that
\begin{equation}
\delta g^{\alpha\beta}\, g_{\alpha\beta} = - \delta g_{\alpha\beta}\, g^{\alpha\beta}.
\end{equation}
Therefore, the varied action $\delta {\cal I}$,
\begin{equation}\label{delta}
\delta {\cal I} = \int d^4 x \sqrt{-g} \left[ \delta g^{\alpha\beta} \left( G_{\alpha\beta} + \kappa T_{\alpha\beta}\right)
+ g^{\alpha\beta} \delta R_{\alpha\beta} \right]=0,
\end{equation}
can be rewritten as
\begin{equation}\label{del}
\delta {\cal I} = \int d^4 x \sqrt{-g} \left[ \delta g^{\alpha\beta} \left( G_{\alpha\beta} - \lambda(r,t) \,g_{\alpha\beta} + \kappa T_{\alpha\beta}\right)\right]=0,
\end{equation}
such that ${{T}}_{\alpha\beta}$ is the background stress tensor
\begin{equation}\label{bt}
{{T}}_{\alpha\beta} =   2 \frac{\delta {{\cal L}_m}}{\delta g^{\alpha\beta}}  - g_{\alpha\beta} {{\cal L}_m}.
\end{equation}
Since the variation of the action is now given by (\ref{del}), then the redefined background Einstein equations take the form
\begin{equation}\label{tr}
\bar{G}_{\alpha \beta} = G_{\alpha \beta} - \lambda(r,t) \,{g}_{\alpha \beta}=-\kappa\, T_{\alpha\beta},
\end{equation}
where now the boundary terms were assimilated in the background Einstein equations, so that
\begin{equation}
\bar{G}^{\alpha \beta}_{\,\,\,\,\,\,;\alpha} = T^{\alpha\beta}_{\,\,\,\,\,\,\,\,;\alpha}=0.
\end{equation}

The expression (\ref{delta}) includes some boundary terms that cannot be arbitrarily neglected to obtain the Einstein's equations. They describe the flux $\delta \Phi$ of $\delta W^{\alpha}=
\delta\Gamma^{\epsilon}_{\beta\epsilon}{g}^{\beta\alpha}-\delta \Gamma^{\alpha}_{\beta\gamma} {g}^{\beta\gamma}$\cite{HE}, through the 3D closed hypersurface $\partial M$:
\begin{equation}\label{fl}
{g}^{\alpha \beta} \delta R_{\alpha \beta}-\delta \Phi =
\left[\delta W^{\alpha}\right]_{|\alpha} - \left(g^{\alpha\beta}\right)_{|\beta}  \,\delta\Gamma^{\epsilon}_{\alpha\epsilon} +
\left(g^{\epsilon\nu}\right)_{|\alpha}  \,\delta\Gamma^{\alpha}_{\epsilon\nu},
\end{equation}
where $"|"$ denotes the covariant derivative on the extended manifold and the flux $\delta \Phi$, is
\begin{equation}
\delta \Phi = \lambda(r,t)\,g^{\alpha\beta}\,\delta g_{\alpha\beta}. \label{flo}
\end{equation}
The variation of the Ricci tensor $\delta R_{\alpha \beta}$, it is calculated on the extended manifold by using the extended Palatini identity\cite{pal,cas}
\begin{equation}
\delta{R}^{\alpha}_{\beta\gamma\alpha}=\delta{R}_{\beta\gamma}= \left(\delta\Gamma^{\alpha}_{\beta\alpha} \right)_{| \gamma} - \left(\delta\Gamma^{\alpha}_{\beta\gamma} \right)_{| \alpha}.
\end{equation}

In order to describe the displacement of the extended manifold: $\delta \Gamma^{\alpha}_{\beta\gamma}$ with respect to the Riemann manifold, driven by the Levi-Civita connections $ \left\{ \begin{array}{cc}  \alpha \, \\ \beta \, \gamma  \end{array} \right\}$, we shall consider the connections given by
\begin{equation}\label{ConexionWeyl}
\Gamma^{\alpha}_{\beta\gamma} = \left\{ \begin{array}{cc}  \alpha \, \\ \beta \, \gamma  \end{array} \right\} + \delta \Gamma^{\alpha}_{\beta\gamma} = \left\{ \begin{array}{cc}  \alpha \, \\ \beta \, \gamma  \end{array} \right\}+ b \,\sigma^{\alpha} g_{\beta\gamma},
\end{equation}
where $\sigma_{\alpha}\equiv \sigma_{,\alpha}$ is the ordinary partial derivative of $\sigma$ with respect to $x^{\alpha}$. On the Riemman manifold it is required that the non-metricity to be null: $\Delta g_{\alpha \beta} = {g}_{\alpha \beta ; \gamma} dx^{\gamma}=0$. However, on the extended manifold the variation of the metric tensor is\footnote{In general, we shall consider variations of arbitrary second range tensors $T_{\alpha\beta}$ as
\begin{displaymath}
\delta T_{\alpha\beta}= T_{\alpha\beta|\epsilon} \,dx^{\epsilon},
\end{displaymath}
so that the variation with respect to the Riemann manifold is
\begin{displaymath}
\frac{\delta T_{\alpha\beta}}{\delta S} = T_{\alpha\beta|\epsilon}\,U^{\epsilon}.
\end{displaymath}
Here, $U^{\epsilon}$ are the components of the velocity for a relativistic observer that moves on the Riemann manifold.}
\begin{equation}
\delta g_{\alpha \beta} = {g}_{\alpha \beta | \gamma} dx^{\gamma} = - b\, (\sigma_{\beta} {g}_{\alpha \gamma} + \sigma_{\alpha} {g}_{\beta \gamma}) dx^{\gamma},
\end{equation}
where ${g}_{\alpha \beta | \gamma}$ is the covariant derivative on the extended manifold generated by the connections (\ref{ConexionWeyl}), with
${g}_{\alpha \beta ; \gamma}=0$.

In order to describe the flux, it is useful to consider $b=1/3$ in (\ref{ConexionWeyl}). In this case we obtain that $\delta W^{\alpha}=-\sigma^{\alpha}$, and hence the expression (\ref{fl}) can be written as
\begin{equation}\label{gauge}
{g}^{\alpha \beta} \delta R_{\alpha \beta}-\delta \Phi =
\left[\delta W^{\alpha}\right]_{|\alpha} - \left(g^{\alpha\beta}\right)_{|\beta}  \,\delta\Gamma^{\epsilon}_{\alpha\epsilon} +
\left(g^{\epsilon\nu}\right)_{|\alpha}  \,\delta\Gamma^{\alpha}_{\epsilon\nu}= \nabla_{\alpha}\delta W^{\alpha}={g}^{\alpha\beta}\left[ \Box
\delta\Psi_{\alpha\beta}-\lambda(r,t)\,\delta g_{\alpha\beta}\right]=-\nabla_{\alpha}\,\sigma^{\alpha}\equiv -\Box \sigma=0,
\end{equation}
where $\delta\Psi_{\beta\gamma}$, can be interpreted as the components of gravitational waves in a more general sense than the standard one. For this reason we will call them space-time waves. They comply with a tensor-wave equation with a source $\lambda(r,t)\delta g_{\alpha\beta}$, but its trace is nonzero: $g^{\alpha\beta} \delta\Psi_{\alpha\beta}\neq 0$. Furthermore, in the standard formalism for gravitational waves, the equation of motion is obtained in a linear perturbative expansion with respect to the background and is valid only as a weak field approximation. The variation of the Ricci tensor on the extended manifold is
\begin{equation}\label{VariacionRicciWeyl}
\delta R_{\alpha \beta}  =  \frac{1}{3} \left[ \nabla_{\beta} \sigma_{\alpha} + \frac{1}{3} \left( \sigma_{\alpha} \sigma_{\beta} + \sigma_{\beta} \sigma_{\alpha} \right) - {g}_{\alpha \beta} \left( \nabla_{\epsilon} \sigma^{\epsilon} + \frac{2}{3} \sigma_{\nu} \sigma^{\nu} \right) \right],
\end{equation}
so that, in agreement with the equations (\ref{gauge}), (\ref{VariacionRicciWeyl}) and (\ref{flo}), we obtain the following equation:
\begin{equation}\label{ein}
\frac{1}{3} \left[ \nabla_{\beta} \sigma_{\alpha} + \frac{1}{3} \left( \sigma_{\alpha} \sigma_{\beta} + \sigma_{\beta} \sigma_{\alpha} \right) - {g}_{\alpha \beta} \left( \nabla_{\epsilon} \sigma^{\epsilon} + \frac{2}{3} \sigma_{\nu} \sigma^{\nu} \right) \right]-\lambda(r,t)\,\delta g_{\alpha\beta} =0.
 \end{equation}
Notice that the transformation (\ref{tr}) preserves the EH action, and the flux that cross the 3D-gaussian hypersurface, $\delta\Phi$, is related to the cosmological parameter $\lambda(r,t)$ and the variation of the scalar field, $\delta\sigma$
\begin{equation}\label{fll}
\delta\Phi = - \frac{2}{3} \lambda(r,t)\,\delta \sigma.
\end{equation}
Moreover, ${\chi}(x^{\epsilon}) \equiv {g}^{\mu\nu} {\chi}_{\mu\nu}$ is a classical scalar field, such that ${\chi}_{\mu\nu}={\delta {\Psi}_{\mu\nu}\over \delta S\,\,\,\,\,\,\,\,}$ describes the space-time waves produced by the evolving source through the 3D-Gaussian hypersurface \begin{equation}
\Box {\chi} = \frac{\delta \Phi}{\delta S}, \label{bb}
\end{equation}
where $\frac{\delta}{\delta S}=U^{\alpha}\frac{d}{dx^{\alpha}}$. Here, $U^{\alpha}=\frac{dx^{\alpha}}{dS}$ are the components of the 4-velocity, given as a solution of the geodesic equation on the Riemann manifold:
\begin{equation}\label{geo}
\frac{dU^{\alpha}}{dS} + \left\{ \begin{array}{cc}  \alpha \, \\ \beta \, \gamma  \end{array} \right\}\,U^{\beta}U^{\gamma} =0,
\end{equation}
with unity squared norm: $U_{\alpha} \,U^{\alpha}=1$. The differential operator $\Box$ that acts on ${\chi}$ in (\ref{bb}), is written in terms of the covariant derivatives defined on the background metric: $\Box\equiv {g}^{\alpha\beta} \nabla_{\alpha}\nabla_{\beta}$, such that $\nabla_{\alpha}$ give us the covariant derivative on the Riemann background manifold, and then (\ref{fll}) will be
\begin{equation}\label{f1}
\frac{d\Phi}{dS} = - \frac{2}{3}\, \lambda(r,t)\,\sigma_{\alpha}\, U^{\alpha},
\end{equation}
where, because of (\ref{gauge}), the scalar field $\sigma$ is the solution of $\Box\sigma=0$.

\section{Collapse with a gravitational attractor: background dynamics}

In order to describe a globally isotropic and homogeneous system which is collapsing due to a gravitational attractor, we shall consider a metric for a isotropic and homogeneous metric, with boundary conditions with a nonzero flux due to a gravitational source that attract the system. The standard metric for a system with scale factor $a(t)$ that collapse without gravitational source in spherical coordinates, is
\begin{equation}\label{me}
g_{\alpha\beta} = {\rm diag} \left[1, -a^2(t), -a^2(t)\,r^2, -a^2(t)\,r^2\,{\rm sin}^2(\theta)\right].
\end{equation}
The mixed components of the Einstein tensor, are
\begin{eqnarray}
G^0_{\hskip.1cm 0}(r,t) & = & -3\,H^2(t), \\
G^1_{\hskip.1cm 1}(r,t) & = & - \left[\frac{\left(2 a \ddot{a}+ \dot{a}^2\right)}{a^2}\right].
\end{eqnarray}
Now, we are interested in add the contribution of a gravitational attractor with variable mass $m(t)$, and a Schwarzschild radius
$r_s(t)=2\,G\,m(t)$. It is assumed that during the collapse, the mass of the black hole is increasing with time, because the BH is dragging matter. Therefore, $\lambda$ will be a function of $r$ and $t$. Therefore, the transformed components of the Einstein tensor, will be
\begin{eqnarray}
\bar{G}^0_{\hskip.1cm 0}(r,t) & = & -3\,H^2(t) - \,\lambda(r,t), \\
 \bar{G}^1_{\hskip.1cm 1}(r,t)& = & - \left[\frac{\left(2 a \ddot{a}+ \dot{a}^2\right)}{a^2}\right]- \,\lambda(r,t).
\end{eqnarray}
For a co-moving observer, the components of the relativistic velocity are
\begin{equation}
U^{0}=1,  \qquad U^{1}=U^{2}=U^3=0,
\end{equation}
and then (\ref{f1}), in this case will be
\begin{equation}\label{f2}
\frac{d\Phi}{dS} = - \frac{2}{3}\, \lambda(r,t)\,\sigma_{0}\, U^{0}.
\end{equation}
The equation of state for a stress tensor described by a perfect fluid
\begin{equation}
T^{\alpha}_{\hskip.1cm\beta} = (P+\rho) U^{\alpha} U_{\beta} - \delta^{\alpha}_{\hskip.1cm \beta}\,P,
\end{equation}
where $\rho$ and $P$ are respectively the energy density and pressure. Therefore, the equation of state: $\omega=P/\rho$, for the system will be
\begin{equation}
\omega(r,t) = -\frac{\bar{G}^1_{\hskip.1cm 1}(r,t)}{\bar{G}^0_{\hskip.1cm 0}(r,t)}.
\end{equation}
In this work we are aimed to study the evolution of a BH with variable (increasing) mass $m(t)$, that is embedded in an collapsing system with scale factor $a(t)$.
A similar system has been studied by McVittie\cite{mcvittie} in the 30' decade, using a background metric to study a BH with fixed mass $m_0$ that is embedded in an
expanding (isotropic and homogeneous) universe, that can be described by a Friedman-Lama\^{\i}tre-Robertson-Walker (FLRW) background metric. However, it is expected that in a collapsing universe, where the collapse is driven by a BH hole that acts as a gravitational attractor, the BH's mass be a dynamical variable that increases as a function of time, rather a fixed mass. Therefore, the Schwarzschild radius must be in this case an increasing function of time $r_s(t)= 2\,G\,m(t)$, and the proper would be to consider the attractor as an external flux of a gravitational field caused by the massive BH. In other words, the collapse must be considered as a consequence of the
BH's existence, and not as an independent fact. The cosmological parameter must be negative, because it is due to an attracting flux of the gravitational field (related to the gravitational potential $V(r,t)=-\frac{r_s(t)}{2\,r\,a(t)}$ in the collapsing system), through the closed 3D-hypersurface. It is known that the cosmological parameter in a spatially flat expanding universe described by a FLRW background metric, is $\lambda(t)=3\,H(t)^2$\cite{jm}. Therefore, for a collapsing system with an gravitational attractor, we shall consider a cosmological parameter $\lambda(r,t)=-3 \,H^2(t)\, {r_s(t)\over a(t) \,r}$, with $H(t)=\dot{a}(t)/a(t)$. In this case one obtains
\begin{eqnarray}
\rho(r,t) &= & \frac{ 3}{8\,\pi\, G} H^2(t)\,\left[1-\left(\frac{r_s(t)}{a\,r}\right)\right], \\
P(r,t) & = & - \frac{3}{8\,\pi\, G} \left[\frac{2\ddot{a} a+ \dot{a}^2\left[1-3\left(\frac{r_s(t)}{a\,r}\right)\right]}{3 \,a^2}\right],
\end{eqnarray}
such that the equation of state for this cosmological parameter is
\begin{equation}
\omega(r,t) = - \frac{1}{3}\left[\frac{2\ddot{a} a +\dot{a}^2\left[1-3\left(\frac{r_s(t)}{a\,r}\right)\right]}{\dot{a}^2\,\left[1-\left(\frac{r_s(t)}{a\,r}\right)\right]}\right].
\end{equation}
Notice that in the limit case where there is no gravitational source, the flux is null $\delta\Phi=0$, and the equation of state assume the expression
\begin{equation}
\lim{\omega(r,t)}_{r_s(t)\rightarrow 0 } =  \rightarrow -\left[\frac{2 a \ddot{a}+\dot{a}^2}{3 \dot{a}^2}\right].
\end{equation}
This limit corresponds to an isotropic and homogeneous universe that collapses. Along our study we
shall require that ${r_s(t)\over a(t)\,r}<1$, in order for describe the exterior to the Schwarzschild radius.

\section{Example: An exponential collapsing model attracted by an increasing massive black-hole}

\begin{figure}
  \centering
    \includegraphics[scale=0.6]{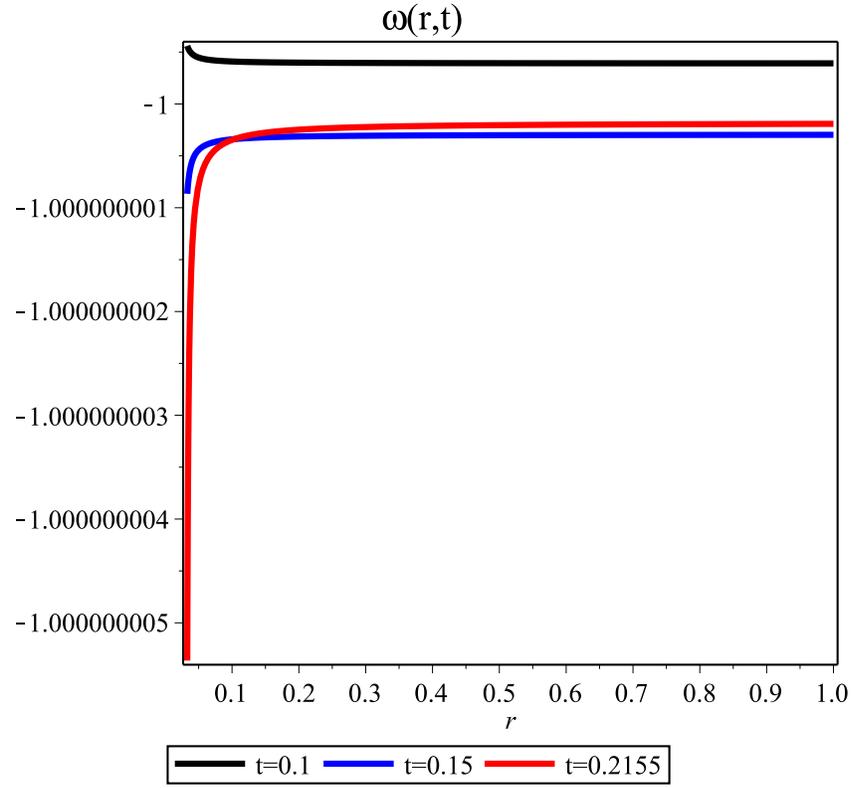}
  \caption{Plot of $\omega$ evaluated at different $t$: $t=0.1$ (black), $t=0.15$ (blue) and $t=0.2155$ (red), as a function of the radius $r$. Along the graphic we have taken $a_0\,H_0=1$, and $m_0\,G=0.01\, a_0$, with $a_0=1$.}
  \label{f1gi121}
\end{figure}

We are aimed to study an exponential collapsing model where $a(t)=a_0\,e^{-H_0\,t}$, such that $a_0> r_s(t)$ is the initial radius of the system and $\dot{a}/a=-H_0<0 $ is a constant. For simplicity we shall consider that the gravitational source has spherical symmetry. In the figure (\ref{f1gi121}) we have plotted $\omega$ during the collapse, as a function of the radius $r$. Notice that as the collapse evolves, $\omega$, which initially takes values up to $-1$, decays to values below $-1$ at the end of the collapse, but reaching the vacuum equation of state for large $r$.

The equation of motion for the scalar field $\sigma$: $\Box\sigma=0$, is given by
\begin{equation}
\ddot{\sigma} + 3 \frac{a}{a} \,\dot{\sigma}-\frac{1}{a^2}\,\nabla^2_{(l=m=0)}\sigma=0,
\end{equation}
with the differential operator
\begin{equation}
\nabla^2_{(l=m=0)} \equiv \left.\frac{1}{r^2} \frac{\partial}{\partial r}\left(r^2 \frac{\partial}{\partial r}\right) + \frac{1}{r^2\sin\theta} \frac{\partial}{\partial \theta} \left(
\sin\theta\,\frac{\partial}{\partial\theta}\right) + \frac{1}{r^2\sin^2\theta} \frac{\partial^2}{\partial\phi^2}\right|_{(l=m=0)}.
\end{equation}
Here, we must set $l=m=0$, because the spherical symmetry of the gravitational source.

\subsection{Solution for $\sigma$}

In order to know the source in $\Box\chi={d\Phi\over dS}$, we must solve the differential equation for $\sigma$:
\begin{equation}
\Box \sigma = 0.
\end{equation}
To make it, we consider a Fourier expansion for $\sigma$, such that the spatial coordinates are spherical
\begin{equation}
\sigma(\vec{r},t) = \int^{\infty}_{0} dk\,k^2\ \, \left[ a_{k00}\,Y_{00}(\theta,\phi)\,{e^{i\,kr}\over r}\,\Sigma_{k0}(t)+ a^{\dagger}_{k00}\,Y^*_{00}(\theta,\phi)\,{e^{-i\,kr}\over r}\,\Sigma^*_{k0}(t)\right],
\end{equation}
where $Y_{00}(\theta,\phi)\equiv Y_{l=0\,m=0}(\theta,\phi)=1$ are the monopolar spherical harmonics. Furthermore, the radial function $R_{\sigma}(kr)={e^{i\,kr}\over r}$, is a
solution of the differential equation
\begin{equation}
\nabla^2_{(l=m=0)}\,R_{\sigma}(kr)+k^2\,R_{\sigma}(kr)=0.
\end{equation}
The solution
\begin{equation}
{e^{i\,kr}\over r}=-ik\,h^{(2)}_0(kr),
\end{equation}
is given by the second kind zeroth order spherical Hankel function: $h^{(2)}_0(kr)$. In general, these functions are defined as $h^{(1,2)}_l(kr)=j_l(kr)\pm i\,y_l(kr)$, which can be written in terms of the spherical Bessel functions $y_l(kr)$ and $j_l(kr)$. Notice that we have set $l=m=0$, because the isotropy and homogeneity is assumed in the source. The coefficients comply with the algebra
\begin{equation}
\left[a_{k00},a^{\dagger}_{k'00}\right] = \delta({k}-{k}')\, ,\hskip .8cm \left[a_{k00},a_{k'00}\right]=\left[a^{\dagger}_{k00},a^{\dagger}_{k'00}\right]=0.
\end{equation}
The differential equations that describes the temporal contribution $\Sigma_{k0}(t)$, is
\begin{equation}
\ddot{\Sigma}_{k0}(t) - 3\,H_0\, \dot{\Sigma}_{k0}(t) + \left(\frac{k}{a_0}\right)^2\,e^{2\,H_0\,t}\,\Sigma_{k0}(t)=0,
\end{equation}
which has the solution
\begin{equation}\label{s1}
\Sigma_{k0}(t)= \left(a_0H_0\right)\,\left[\left(\frac{k}{a_0H_0}\right)\,e^{H_0\,t} + i\right]\,e^{i\left(\frac{k}{a_0H_0}\right)e^{H_0t}},
\end{equation}
with $k_0(t)\geq k\geq a_0H_0$, in order to recover the wavelengths biggest than the initial Schwarzschild radius and smallest than initial Hubble horizon, which is related with initial size of the collapsing system. The time dependent function $k_0(t)$ is:
\begin{equation}\label{ko}
k_0(t)= \frac{3}{2}\,\left(H_0a_0\right)\,e^{-H_0 \,t}.
\end{equation}
Therefore, the Fourier expansion for $\sigma$, will be
\begin{equation}
\sigma(\vec{r},t) = \frac{1}{r}\,\int^{k_0(t)}_{a(t)\, H_0} dk\, k^2\,\left[ a_{k00}\,Y_{00}(0,\phi)\,e^{i\,kr}\,\xi_{k0}(t)+ a^{\dagger}_{k00}\,Y^*_{00}(0,\phi)\,e^{-i\,kr}\,\xi^*_{k0}(t)\right],
\end{equation}
for $0\leq t\leq t_*$, such that $t_*$ is the time for what $k_0(t_*)=H_0\,a(t_*)$.

\subsection{Solution for $\chi$}

In order to know the space-time waves emitted during the collapse, we must solve the equation for $\chi$ with the source (\ref{fl})
\begin{equation}
\Box \chi=- \frac{2}{3}\, \lambda(r,t)\,\sigma_{\alpha}\, U^{\alpha}.
\end{equation}
In our example $U^{\alpha}=(1,0,0,0)$, $\sigma_{0}\equiv \dot\sigma$ and the cosmological parameter takes into account the flux of a gravitational attractor
$\lambda(r,t)=-3 \,H^2(t)\, {r_s(t)\over a(t) \,r}$. The Fourier expansion for $\chi$, must take into account, in principle, all the harmonics in spherical coordinates, once the source is originated by the flux of
$\sigma^{\alpha}$
\begin{equation}
\chi(\vec{r},t) = \sum_{L,M=-L,..,L}\,\int^{\infty}_{0} dK\,K^2\ \, \left[ b_{KLM}\,Y_{LM}(\theta,\phi)\,R_{\chi}(Kr)\,\xi_{KL}(t)+ b^{\dagger}_{KLM}\,Y^*_{LM}(\theta,\phi)\,R^*_{\chi}(Kr)\,\xi^*_{KL}(t)\right],
\end{equation}
with the following algebra for the coefficients
\begin{equation}
\left[b_{KL0},b^{\dagger}_{K'L'0}\right] = \delta({K}-{K}')\,\delta_{LL'}\, ,\hskip .8cm \left[b_{KL0},b_{K'L'0}\right]=\left[b^{\dagger}_{KL0},b^{\dagger}_{K'L'0}\right]=0.
\end{equation}

\begin{figure}
  \centering
    \includegraphics[scale=0.6]{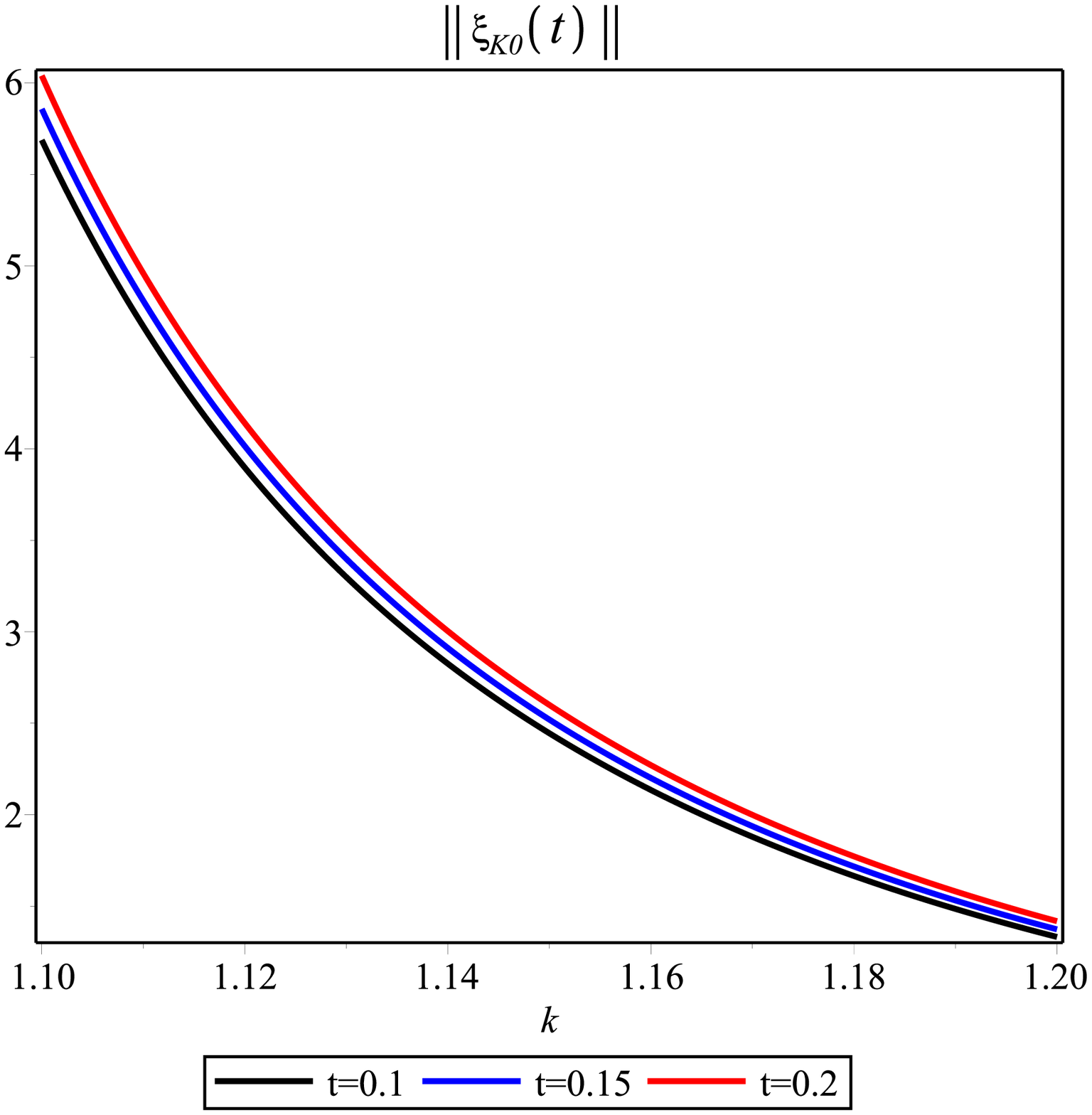}
  \caption{Plots of $\parallel \xi_{K0}(t)\parallel$ evaluated at different $t$: $t=0.1$ (black), $t=0.15$ (blue) and $t=0.2$ (red), as a function of the wavenumber $K$. Along the graphic we have taken $a_0\,H_0=1$, and $m_0\,G=0.01\, a_0$, with $a_0=1$.}
  \label{f2gi121}
\end{figure}

Here, the explicit differential equation for $\chi$ is
\begin{equation}\label{ch}
\ddot{\chi} + 3 H(t) \,\dot{\chi}-\frac{1}{a^2}\,\nabla^2_{(L=1,M=0)}\chi=- \frac{2}{3}\, \lambda(r,t)\,\sigma_{\alpha}\, U^{\alpha},
\end{equation}
with the differential operator
\begin{equation}
\nabla^2_{(L=1,M=1,0,-1)} \equiv \left.\frac{1}{r^2} \frac{\partial}{\partial r}\left(r^2 \frac{\partial}{\partial r}\right) + \frac{1}{r^2\sin\theta} \frac{\partial}{\partial \theta} \left(
\sin\theta\,\frac{\partial}{\partial\theta}\right) + \frac{1}{r^2\sin^2\theta} \frac{\partial^2}{\partial\phi^2}\right|_{(L=1,M=1,0,-1)}.
\end{equation}
The term with $L=1$ comes because the gravitational source described by $\lambda(r,t)$ is proportional to $1/r$, so that we must include a solution with a superior order in the radial contribution, which is described by the spherical Bessel functions, when we make variables separation in the differential equation(\ref{ch}). The solution for our problem gives the spherical harmonics with $L=1$ and $M=1,0,-1$: $Y_{LM}(\theta,\phi)=\sqrt {(2L+1)\frac{(L-M)!}{(L+M)!}}\,P_{L }^{M}(\cos {\theta })\,e^{i\,M\,\phi}$, where $P_{L}^{M}(\cos {\theta })$  are the Legendre polynomial. The radial functions $R_{\chi}(Kr)$, in general are given by linear combination of the spherical Hankel functions: $h^{(1,2)}_L(Kr)=j_L(Kr)\pm i\,y_L(Kr)$. If we impose that $b_{K10}=b_{K00}=a_{k00}$ and $b_{K11}=b_{K1-1}=0$, the solution for our particular problem is
\begin{equation}
R_{\chi}(Kr) = h^{(1)}_0(Kr)+ i\,h^{(1)}_1(Kr)\equiv \frac{e^{i\,Kr}}{(K\,r)^2},
\end{equation}
where always we can set $\theta=0$, because the normal on the closed hypersurface is parallel to the radial unitary vector $\hat{r}$. The Legendre polynomial with $L=1$, $M=0$, evaluated at $\theta=0$, is $\left.P_{1 }^{0}(\cos {\theta })\right|_{\theta=0}={1\over 2} \sqrt{{3\over \pi}}$. Finally, the differential equations that describes the temporal contribution $\xi_{K0}(t)$, is
\begin{equation}
\ddot{\xi}_{K0}(t) - 3\,H_0 \dot{\xi}_{K0}(t) + \frac{K^2}{a_0^2}\,e^{2H_0\,t}\xi_{K0}(t)=4\sqrt{\frac{\pi}{3}}\,H_0^2\,\frac{r_s(t)}{a_0}\,e^{H_0\,t}\,e^{i\,(k-K)}\,\dot{\Sigma}_{k0}(t).
\end{equation}
To solve this equation we must use the solution of the source (\ref{s1}). For example, for the special case where the mass of the source increases exponentially during the collapse: $m(t)=m_0\,e^{H_0t}$, we obtain the general solution
\begin{eqnarray}
\xi_{K0}(t) &=& \,\left(a_0H_0\right)\,\left[\left(\frac{K}{a_0H_0}\right)\,e^{H_0\,t} + i\right]\,e^{i\left(\frac{K}{a_0H_0}\right)e^{H_0t}} +\,\frac{8}{3} \left[\frac{\sqrt{3\,\pi} H^2_0 G\,m_0\,k^2}{\left(K-k\right)^2\left(K+k\right)^2}\right] \,e^{i\left(\frac{k}{a_0H_0}\right)e^{H_0t}}\,e^{-i(K-k)}\nonumber \\
& \times & \left[2\,i\,H^2_0\,a^2_0+i\,(K^2-k^2)e^{2\,H_0\,t}+2H_0a_0k\,e^{H_0\,t}\right], \label{ss}
\end{eqnarray}
where the instantaneous Schwarzschild radius is $r_s(t)= 2\,G\,m(t)$ and $K>k$. However, it is expected that in absence of gravitational sources there are no space-time waves, so that the physical solution that describes the space-time waves is only given by the particular contribution of (\ref{ss}):
\begin{eqnarray}
\xi_{K0}(t) = \,\frac{8}{3} \left[\frac{\sqrt{3\,\pi} H^2_0 G\,m_0\,k^2}{\left(K-k\right)^2\left(K+k\right)^2}\right] \,e^{i\left(\frac{k}{a_0H_0}\right)e^{H_0t}}\,e^{-i(K-k)}
\, \left[2\,i\,H^2_0\,a^2_0+i\,(K^2-k^2)e^{2\,H_0\,t}+2H_0a_0k\,e^{H_0\,t}\right]. \label{ss1}
\end{eqnarray}
Finally, the Fourier expansion for $\chi$, will
be
\begin{equation}
\chi(\vec{r},t) = \frac{1}{r^2}\,\int^{k_0(t)}_{a(t)\, H_0} dK\, \,\left[ b_{K10}\,Y_{10}(0,\phi)\,e^{i\,Kr}\,\xi_{K0}(t)+ b^{\dagger}_{K10}\,Y^*_{10}(0,\phi)\,e^{-i\,Kr}\,\xi^*_{K0}(t)\right],
\end{equation}
with $b_{K10}=b_{K00}=a_{k00}$. Notice that the origin of the space-time waves is the temporal variation of the mass, which drives
an increment in the Schwarzschild radius that alters the geometrical environment of the BH, and generates space-time waves. Therefore,
we have introduced the limits in the integral that run from the minimum wavenumber $K_{min}=a_0H_0$, related to the initial size of the ball which collapses, and the dynamical maximum wavenumber $k_0(t)=\frac{3}{2} a_0 H_0\,e^{-H_0 t} \geq K_{min}$, related to the size of the Schwarzschild horizon that increases with time as $r_s(t)=2m_0\,G\, e^{H_0 t}$. Therefore, the time to reach this maximum will be:
\begin{equation}
t_* = -\frac{1}{2H_0}\,{\rm ln}\left[\frac{2}{3}\right] >0.
\end{equation}
In the figure (\ref{f2gi121}) we have plotted the norm $\left \|\xi_{K0}(t)\right\|$ of the temporal modes for $\chi$, for different instants : $t=0.1$ (black), $t=0.15$ (blue) and $t=0.2$ (red), as a function of the wavenumber $K$. Notice that the magnitude of the mode increases with time as the collapse evolves and therefore the front of the
space-time wave becomes more important.

\section{Final Comments}

We have studied the dynamics of a gravitational collapse in an asymptotic homogeneous and isotropic universe in the framework of RQG for a flux trough a 3D-closed
hypersurface which has the form: $\delta R_{\alpha\beta} = \lambda(r,t)\,\delta g_{\alpha\beta}$, when the gravitational source increases with time during the collapse of the system. In particular, we have explored an exponential collapse and the space-time waves emitted during this collapse due to the back-reaction effects described by
$\sigma$, which has a geometrical origin. Its derivative, $\sigma^{\alpha}$ is the responsible of the generation for the extended manifold: $\delta \Gamma^{\alpha}_{\beta\epsilon}=\frac{1}{3}\,\sigma^{\alpha}\,g_{\beta\epsilon}$, on which is described the dynamics of these waves. In other words, $\sigma$ is the geometrical source (more precisely, its time derivative), of the space-time waves produced during the collapse. Notice that the equation (\ref{bb}), can be written as
\begin{displaymath}
\Box {\chi} =  \frac{\delta \Phi}{\delta S}\equiv \frac{\delta R}{\delta S}=\lambda(r,t)\,g^{\alpha\beta}\frac{\delta g_{\alpha\beta}}{\delta S}
=2\,\lambda(r,t)\,\frac{1}{V}\frac{\delta V}{\delta S}=- \frac{2}{3}\, \lambda(r,t)\,\sigma_{\alpha}\, U^{\alpha},
\end{displaymath}
where $V\equiv \sqrt{-g}$ is the volume of the manifold. Notice that the last equality is $ \frac{\delta \Phi}{\delta S}=- \frac{2}{3}\, \lambda(r,t)\,\dot\sigma$ when the relativistic observer is comoving with the collapse. Therefore, for the family of cases here studied, with $\delta R_{\beta\epsilon}=\lambda(r,t)\,\delta g_{\beta\epsilon}$ the source of space-time waves has a geometrical origin, but also a physical one, because it depends on $\lambda(r,t)$, which is considered in this work as originated by a gravitational source. We must recall that $\delta g_{\alpha\beta} \neq 0$, because the extended manifold has nonzero non-metricity.

\section*{Acknowledgements}

\noindent This research was supported by the CONACyT-UDG Network Project No. 294625 "Agujeros Negros y Ondas Gravitatorias". M. B. acknowledges CONICET, Argentina (PIP 11220150100072CO), and UNMdP (EXA955/20) for financial support. J. M. H. acknowledges CONACYT scholarship.

\newpage



\bigskip


\begin{thebibliography}{99}
\bibitem{1} J. Yadav, S. Bharadwaj, B. Pandey, T. R. Seshadri, Mon. Not. Roy. Astron. Soc. \textbf{364}: 601 (2005).
\bibitem{op} J. R. Oppenheimer, H. Snyder, Phys. Rev. \textbf{56}: 455 (1939).
\bibitem{t} R. C. Tolman, Proc. Natl. Acad. Sci. USA \textbf{20}: 169 (1934).
\bibitem{b} H. Bondi, Mon. Not Astron. Soc. \textbf{107}:410 (1947).
\bibitem{pen} R. Penrose, Riv. Nuovo Cim. \textbf{1}: 252 (1969).
\bibitem{w} D. Christodoulou, Ann. Maths. \textbf{149}: 183 (1999).
\bibitem{s} M. Dafermos, I. Rodnianski, Clay Math. Proc. \textbf{17}: 97 (2013).
\bibitem{Sharma} R. Sharma, S. Das, R. Tikekar, Gen. Relat. Grav. {\bf 47}: 25 (2015).
\bibitem{HS} L. Herrera, N. O. Santos, Phys. Rev. {\bf D70}: 084004 (2004).
\bibitem{B} M. Bojowald, R. Goswami, R. Maartens and P. Singh, Phys. Rev. Lett. {\bf 95}: 091302 (2005).
\bibitem{J} R. Goswami, P. S. Joshi, P. Singh, Phys. Rev. Lett. {\bf 96}: 031302 (2006).
\bibitem{mal} D. Malafarina, B. Toshmatov, N. Dadhich, Phys. Dark Univ. \textbf{30}: 100598 (2020).
\bibitem{pul} F. Ben\'{\i}tez, R. Gambini, L. Lehner, S. Liebling, J. Pullin, Phys. Rev. Lett. \textbf{124}: 071301 (2020).
\bibitem{Gundlach} C. Gundlach, Living Rev. Rel. {\bf 2}: 4 (1999).
\bibitem{GJ} R. Goswami, P. S. Joshi, Phys. Rev. {\bf D65}: 027502 (2004).
\bibitem{G} R. Giambo, Class. Quant. Grav. {\bf 22}: 2295 (2005).
\bibitem{jmc} J. Mendoza Hern\'andez, M. Bellini, C. Moreno, Phys. Dark Univ. \textbf{26}: 100395 (2019).
\bibitem{jmc1} J. Mendoza Hern\'andez, M. Bellini, C. Moreno, Phys. Dark Univ. \textbf{23}: 100251 (2019).
\bibitem{HE} S. W. Hawking, G. F. R. Ellis. {\em The large scale structure of space-time}. Cambridge Monographs on Mathematical Physics.
Cambridge University Press. Cambridge, UK (1973).
\bibitem{pal} A. Palatini. {\it Deduzione invariantiva delle equazioni gravitazionali dal principio di Hamilton}, Rend. Circ. Mat. Palermo {\bf 43}: 203-212  (1919). [English translation by R. Hojman and C. Mukku in P. G. Bergmann and V. De Sabbata (eds.) Cosmology and Gravitation, Plenum Press, New York (1980)].
\bibitem{cas} A. Guarnizo, L. Casta\~neda, J. M. Tejeiro, Gen. Rel. Grav. \textbf{42}: 2713 (2010).
\bibitem{mcvittie} G. C. McVittie, MNRAS \textbf{93}: 325 (1933).
\bibitem{jm} J. I. Musmarra, M. Bellini, Phys. Dark Univ. \textbf{30}: 100670 (2020).
\end{thebibliography}
\end{document}